\documentclass{PoS}

\usepackage{xspace}
\usepackage{amsmath}
\usepackage{lineno}

\def \lsi {LS~I~+61$^\circ$~303\xspace}
\def \hess {HESS~J0632+057\xspace}

\title{Looking for Gamma-ray Emission\\ from TeV Binary Candidates with HAWC}

\ShortTitle{HAWC Binary Detection}

\author{\speaker{Segev BenZvi}$^a$ for the HAWC Collaboration$^b$\\
\llap{$^a$}Department of Physics and Astronomy, University of Rochester, Rochester, NY, USA \\
\llap{$^b$}For a complete author list, see \href{http://www.hawc-observatory.org/collaboration/icrc2015.php}{www.hawc-observatory.org/collaboration/icrc2015.php}. \\
Email: \email{sybenzvi@pas.rochester.edu}}

\abstract{

  The Milky Way contains hundreds of binary systems which are known to emit in
  radio and X-rays, but only a handful of binaries have been observed to
  produce very high-energy gamma rays.  In addition, the emission mechanisms
  which produce the gamma rays in the few known sources are not well
  understood.  To improve the statistics of binary sources in the TeV band, the
  High-Altitude Water Cherenkov Gamma-ray Observatory, or HAWC, has begun to
  carry out a simultaneous survey of TeV binary candidates in the Northern
  Hemisphere between $100$~GeV and $100$~TeV.  HAWC is a surface array that
  records air showers from cosmic rays and gamma rays with a high uptime and
  wide field of view, making it well-suited to observe time-dependent emission
  from objects such as TeV binaries.  We describe the sensitivity of HAWC to
  periodic emission from Galactic sources of gamma rays and present data from
  the first year of observations with the partially constructed observatory.

}

\FullConference{The 34th International Cosmic Ray Conference,\\
		30 July- 6 August, 2015\\
		The Hague, The Netherlands}

\begin{document}

\section{Introduction}

Gamma-ray binaries are compact Galactic objects such as pulsars or black holes
which orbit massive stars.  It is thought that gamma rays are produced in these
systems when relativistic particles accelerated by a pulsar encounter the
envelope of the companion star, or when the compact object accretes material
from its companion and produces jets of relativistic particles
\cite{Mirabel:2012ye}. The transient behavior in these systems, which includes
both flaring and periodic emission, makes them ideal for multiwavelength
observations and modeling. Multi-messenger observations are also a possibility;
some gamma-ray binaries may accelerate hadronic cosmic rays and produce
transient emission of neutrinos \cite{Abbasi:2011ke}.

To date, only five gamma-ray binaries have been observed at TeV.
Three of these objects can be seen from the Northern Hemisphere: \lsi
\cite{Acciari:2008hg,Acciari:2011is}, \hess
\cite{Aharonian:2007nh,MAGIC:2012aa}, and LS~5039 \cite{Aharonian:2005eb,
Aharonian:2006vu}.  Thus far an unbiased survey of TeV binary candidates has
not been attempted; all measurements have been pointed observations made with
imaging air Cherenkov telescopes (IACTs).  Moreover, the TeV emission from
compact binary candidates has proven difficult to predict using observations
made at lower energies.  For example, the intense radio and X-ray source
Cygnus~X-3 has been observed at GeV but not at TeV despite extensive
observational campaigns \cite{Archambault:2013yva}.  Meanwhile, the TeV source
\hess has not been observed at GeV \cite{Caliandro:2013oda}.

The low statistics of the current population of known TeV binaries, and the
unexplained mismatches between emission in the X-ray, GeV, and TeV bands
motivate a long-term search for new TeV binary systems.  Due to its high uptime
and wide field of view, the High Altitude Water Cherenkov Observatory (HAWC) is
an excellent instrument to carry out an unbiased survey of binary systems in
the northern sky.

\section{The HAWC Observatory}

HAWC is a gamma-ray and cosmic ray detector located $4100$~m above sea level in
Sierra Negra, Mexico.  The detector is an air shower array comprising 300
close-packed water Cherenkov detectors (WCDs).  Each WCD is a steel tank lined
with a light-tight polypropylene bladder and filled with $200$~kL of purified
water. Four hemispherical photomultiplier tubes (PMTs) at the bottom of each
WCD, for a total of 1200 PMTs in the full array, observe the Cherenkov light
produced when air shower particles pass through the detector.

By combining the timing information and spatial pattern of PMTs triggered by
an air shower, it is possible to reconstruct the arrival direction and
identify the particle type of the air shower primary.  Using simple topological
cuts we can discriminate the air showers produced by hadronic cosmic rays from
the air showers produced by gamma rays and suppress the $20$~kHz all-sky
hadronic background.  Details on the event reconstruction and gamma-hadron
discrimination techniques used in HAWC are given in \cite{Smith:2015}.

The instantaneous field of view of HAWC is about $2$~sr, and the uptime of the
observatory is $>90\%$.  While HAWC is less sensitive to pointlike sources of
gamma rays than the current generation of IACTs, the observatory can be used to
record emission from a large number of sources simultaneously.  This makes HAWC
well-suited for long-term observations of variable sources, including the
period-modulated and flaring emission from $\gamma$-ray binaries.

\section{Sensitivity of HAWC to Periodic and Flaring Emission}

\begin{figure}[ht]
  \includegraphics[width=\textwidth]{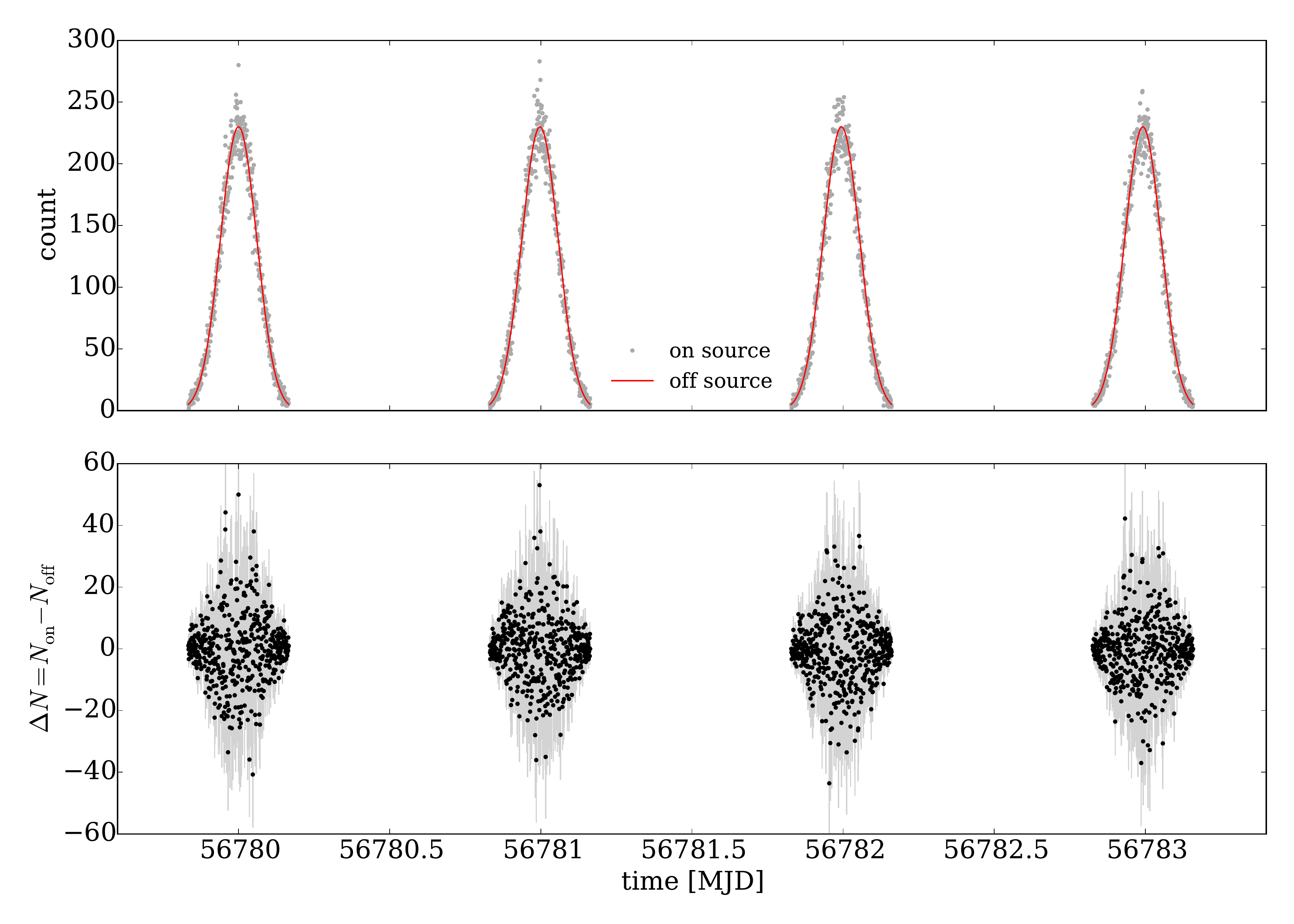}
  \caption{\label{fig:fake_transit}
  {\sl Top}: Simulated transit of a source %at declination $\delta=19^\circ$
  observed with the HAWC detector.
  {\sl Bottom}: difference between observed counts in the search region around
  the source (``on-source'') and expected counts estimated using an area
  outside the search region (``off source''). The gray lines indicate the
  uncertainties in the residual counts.}
\end{figure}

\begin{figure}[ht]
  \includegraphics[width=\textwidth]{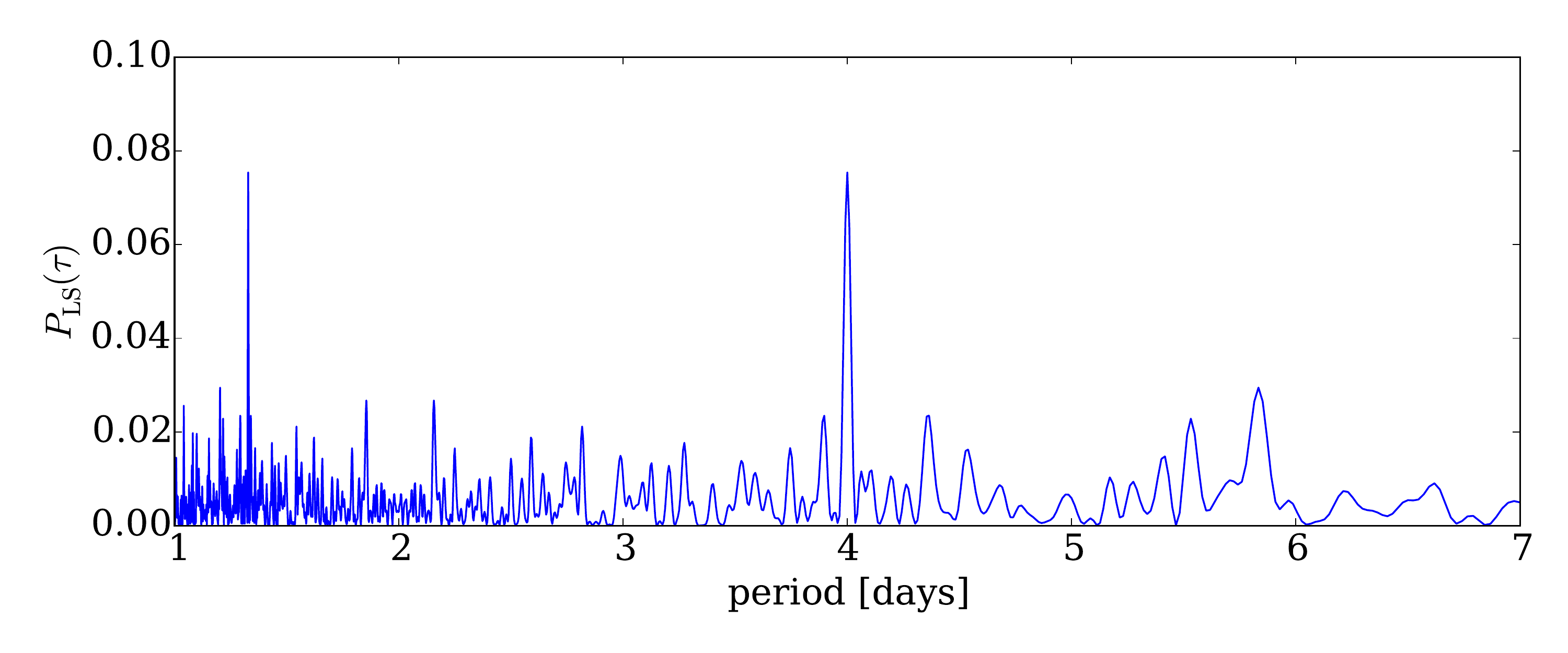}
  \includegraphics[width=\textwidth]{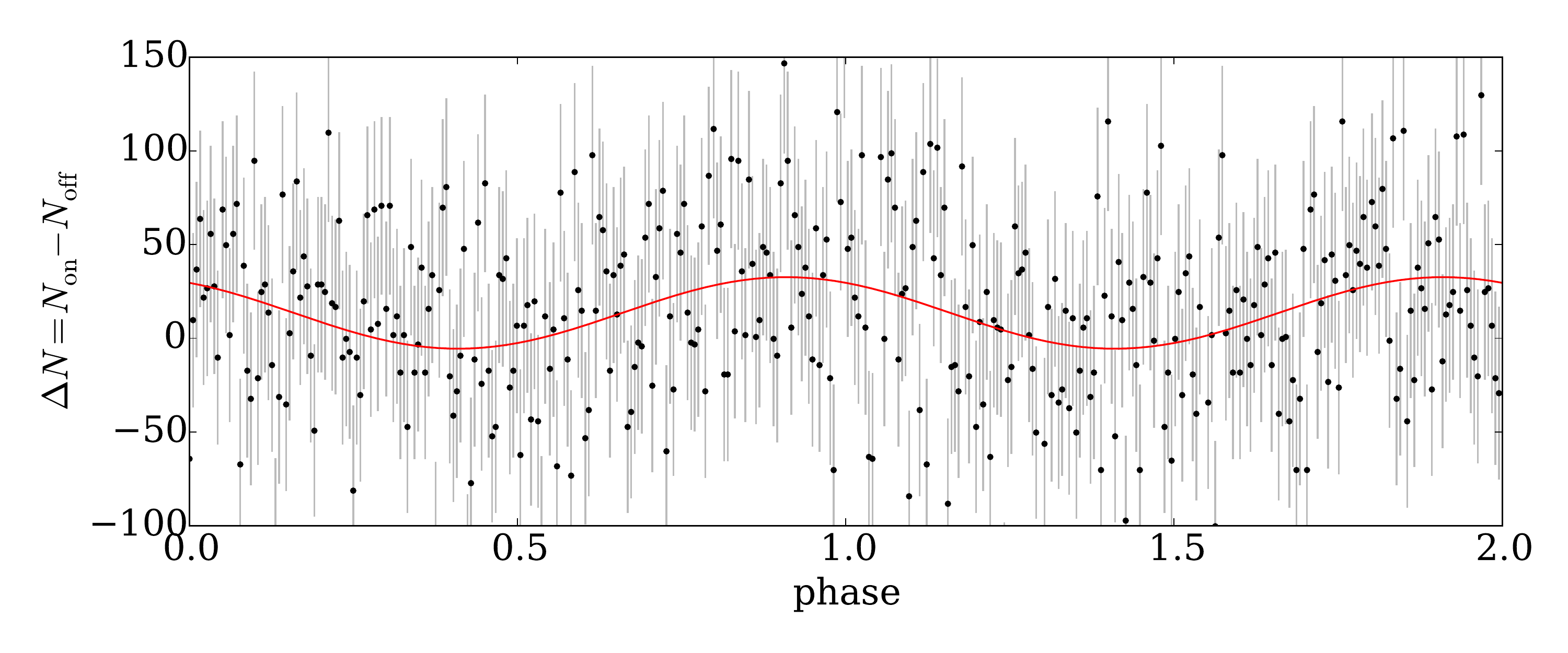}
  \caption{\label{fig:lspower}
  {\sl Top}: Lomb-Scargle periodogram calculated for a year of simulated data 
  from a source which transits $15^\circ$ from the detector zenith. The
  source was modeled to have $5\%$ of the flux of the Crab above $1$~TeV and
  a $4$-day periodicity.
  {\sl Bottom}: Period-folded 12-month light curve from the simulated source,
  expressed as the difference $\Delta N$ between on- and off-source counts.
  The line shows the best-fit signal corresponding to a 4-day period.}
\end{figure}

Four simulated transits of a point source of gamma rays through the field of
view of HAWC are shown in Fig.~\ref{fig:fake_transit}.  The arrival times of
the events are plotted using one-minute time bins.  The plot depicts the total
number of events observed in 9 bins in $N_\text{hit}$, or the number of hits
observed per event, with angular and gamma-hadron cuts optimized in each
$N_\text{hit}$ bin for a point source of gamma rays \cite{Abeysekara:2013tza}.
A typical transit, defined as the period in which a source is above $45^\circ$
elevation in the field of view of the detector, lasts six hours.

HAWC uses two approaches to study point sources.  The first is a basic
cut-and-count analysis, in which the estimated number of cosmic ray events
($N_\text{off}$, shown as the solid line in the top panel of
Fig.~\ref{fig:fake_transit}) is subtracted from the total number of events
observed ($N_\text{on}$, the points in the top panel of
Fig.~\ref{fig:fake_transit}).  The resulting residual counts $\Delta
N=N_\text{on}-N_\text{off}$, shown in the bottom panel of
Fig.~\ref{fig:fake_transit}, can be studied for time dependent behavior.  This
is the approach used in the study of flares in the HAWC online monitoring
system \cite{Weisgarber:2015}.  A second approach, which involves a parametric
maximum-likelihood fit of the gamma-ray spectrum and source location, can also
be used to produce a light curve of the integral flux above some energy
threshold \cite{Younk:2015, Lauer:2015}. The cut-and-count approach has the
advantage of model independence, while the maximum likelihood analysis provides
better sensitivity and a simple interpretation of data in terms of physical
fluxes.

\subsection{Periodic Emission}

To detect period-modulated emission in a gamma-ray source, we estimate the
Lomb-Scargle periodogram of the residual counts $\Delta N$.  The periodogram of
a year of simulated transits similar to the source shown in
Fig.~\ref{fig:fake_transit} is plotted in the top panel of
Fig.~\ref{fig:lspower}.  The emission from the simulated source is modeled as a
sinusoid with a period of $4$~days.  The source transits $15^\circ$ from the
detector zenith and has $5\%$ of the flux of the Crab Nebula above $1$~TeV in
its high state.  Defining the high state as the time when the signal is within
10\% of its maximum value, the duty cycle of the source is 20\%.  This flux is
typical of the emission seen from the binary systems in the northern hemisphere
\cite{Aharonian:2007nh,MAGIC:2012aa,Aharonian:2005eb,Aharonian:2006vu}.  The
period-folded residual counts $\Delta N$ for 365 simulated transits is shown in
the bottom panel of Fig.~\ref{fig:lspower}.

The periodogram contains a peak at $4$~days corresponding to the simulated
periodicity of the source.  Additional peaks observed at higher frequencies are
higher harmonics and artifacts of the observation window used to simulate the
light curve.  To estimate the significance of the detection of a periodic
signal, we compare the height of the maximum in the periodogram at $4$~days to
a distribution of maxima from data sets with no sinusoidal modulation present.
Using the distribution of maxima in the background-only data sets, we calculate
a $p$-value.

Given the current sensitivity of the detector \cite{Abeysekara:2013tza}, we
estimate that with this analysis it will take approximately 1000~transits, or
3~years, to observe periodicity at the $5\sigma$ level in this kind of source.
Note that this calculation is based on a conservative and not particularly
well-optimized estimate of the gamma-hadron discrimination power of the
detector.  The current sensitivity of the detector has achieved a $Q$-value of
approximately $5$, where $Q$ is defined as the ratio of the gamma-ray shower
selection efficiency $\epsilon_\gamma$ to the cosmic-ray selection efficiency
$\epsilon_\text{CR}$,
\[
  Q=\epsilon_\gamma/\sqrt{\epsilon_\text{CR}}.
\]
Recent efforts to improve the background suppression power of the
point source analysis indicate that substantial improvements to $Q$ are
possible \cite{Hampel:2015,Capistran:2015}.  An improvement to $Q\approx10$,
which can be achieved with new gamma-hadron separators, would result in a
$5\sigma$ detection of the source in Fig.~\ref{fig:lspower} in 12
months.

\subsection{Observation of Flares}

Gamma-ray binaries are also known to produce significant flares.  For example,
the binary system \lsi has produced flaring emission orders of magnitude above
its quiescent flux lasting for several weeks \cite{Holder:2014}.  While \lsi
transits at an elevation of $48^\circ$ in HAWC, where the sensitivity of the
detector is poor, a similar source which flares closer to the detector zenith
should be easily detected.  For example, a binary system which flares at $20\%$
of the Crab flux would be observed at $5\sigma$ within a week, given a
conservative estimate of the sensitivity of HAWC \cite{Abeysekara:2013tza}.
The strength and duration of this flare would be large but not unprecedented
for a binary system.
%; such a flare was observed in \lsi in late 2014
%\cite{Holder:2014}.
An extreme flare with the same integral flux as the Crab
Nebula above $1$~TeV would be observed in less than one transit.

The HAWC Collaboration is running an online monitoring program to observe
flares from blazars in real time.  Included in the monitoring program are 30
Galactic TeV binary candidates.  Flares from these objects can be used to
trigger follow-up observations with sensitive IACTs.  Details of the online
monitoring program are available in \cite{Weisgarber:2015}.

\section{First Data from HAWC}

The construction of the HAWC tanks was completed in December 2014, but the
detector was operated in several configurations during construction.  Prior to
September 2014, the detector operated for approximately one year with 106 to
133 live WCDs, a configuration called HAWC-111.  During Fall 2014, most of the
remaining WCDs were commissioned, with HAWC-250 operations beginning on
November 26, 2014.  The event rate in the detector during this period ranged
between 10~kHz and 15~kHz, with 99.9\% of events coming from cosmic-ray air
showers.

\begin{figure}[ht]
  \includegraphics[width=\textwidth]{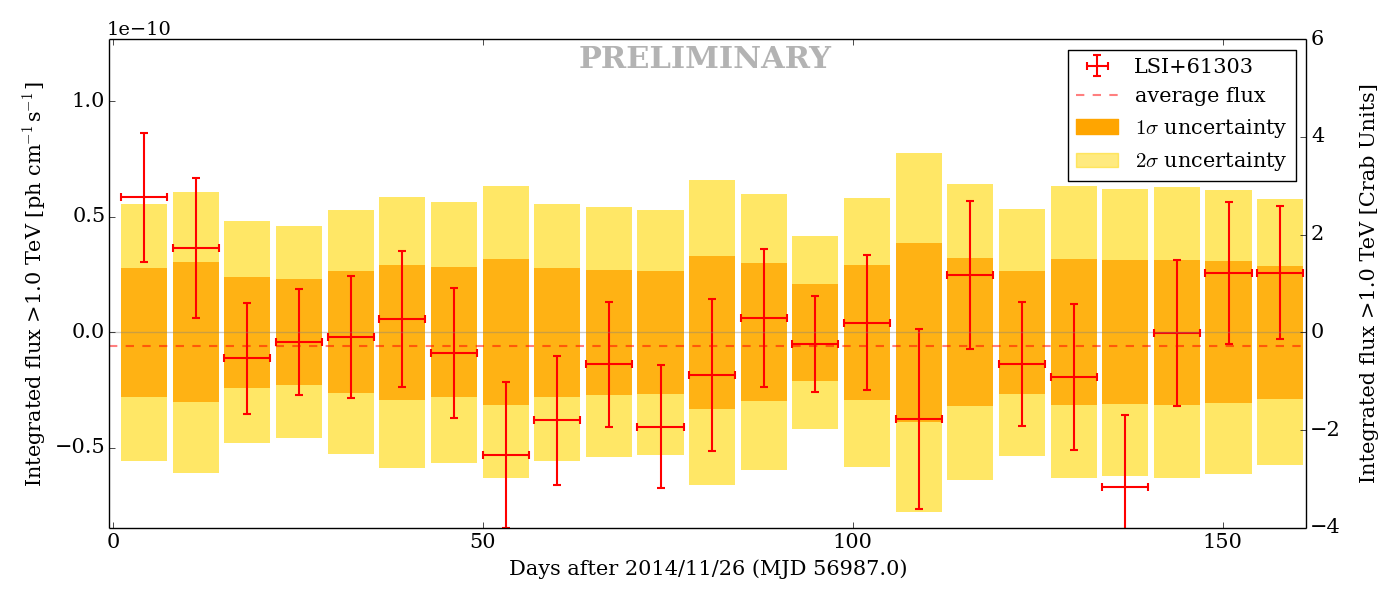}
  \caption{\label{fig:lsi_curve} Integral gamma-ray fluxes (red
  crosses) calculated in 7-day bins from the position of \lsi recorded after
  November 26, 2014 with HAWC-250.  The uncertainties in the flux are large due
  to the poor statistics from this high declination source.  The orange and
  yellow bands indicate the $68\%$ and $95\%$ intervals of the background
  expected at this location in each 7-day bin.  The average flux measured from
  \lsi during this period is shown by the dashed line.}
\end{figure}

As expected, the sensitivity of the HAWC-111 detector and the duration of the
current HAWC-250 data were not sufficient to observe periodicity in event rate
from the locations of the known binary systems in the Northern Hemisphere. 

In October and November 2014 the binary system \lsi exhibited a significant
flare \cite{Holder:2014}.  While the HAWC detector suffered downtime in late
October and early November due to the commissioning of HAWC-250, the
observatory operated continuously after November 26 and was able to monitor
gamma rays from the location of \lsi.  The light curve is shown in
Fig.~\ref{fig:lsi_curve}, expressed in units of the integral flux from the Crab
Nebula.  The sensitivity of the detector is poor for high-declination sources
(two orders of magnitude lower than at the detector zenith).  As a result, no
excess counts or statistically significant variability were observed from \lsi
during this period.

\section{Conclusion}

Due to its high uptime and large field of view, HAWC is a good detector to
monitor periodic and flaring emission from Galactic gamma-ray sources such as
binary systems.  Using a reasonable estimate of the gamma-hadron separation
power of the full detector, we expect to find statistically significant
periodicity in the emission from a binary source like LS~5039 (3.9~day orbital
period) within the first year of detector operations.  The binary system \hess
has a hard spectrum and is located at a favorable declination for HAWC, so it
should also be observed within roughly one year.  However, it has a relatively
long orbital period (315~days) that will take several years to measure
completely.

A set of 30 TeV binary candidates, mostly short-period X-ray binaries, is being
monitored as part of the HAWC Online Flare Monitor.  Flares from these objects,
when observed, will be reported to IACTs for sensitive follow-up observations.

\section*{Acknowledgments}
\footnotesize{
We acknowledge the support from: the US National Science Foundation (NSF);
the US Department of Energy Office of High-Energy Physics;
the Laboratory Directed Research and Development (LDRD) program of
Los Alamos National Laboratory; Consejo Nacional de Ciencia y Tecnolog\'{\i}a
(CONACyT),
Mexico (grants 260378, 55155, 105666, 122331, 132197, 167281, 167733);
Red de F\'{\i}sica de Altas Energ\'{\i}as, Mexico;
DGAPA-UNAM (grants IG100414-3, IN108713,  IN121309, IN115409, IN111315);
VIEP-BUAP (grant 161-EXC-2011);
the University of Wisconsin Alumni Research Foundation;
the Institute of Geophysics, Planetary Physics, and Signatures at Los Alamos
National Laboratory;
the Luc Binette Foundation UNAM Postdoctoral Fellowship program.
}

\bibliography{icrc2015-0217}

\end{document}